\documentclass[pra,a4paper,aps,twocolumn,showpacs,superscriptaddress,groupedaddress]{revtex4}
\usepackage[dvips]{graphicx}
\usepackage{ae}
\usepackage[T1]{fontenc}
\usepackage[ansinew]{inputenc}
\usepackage{amsmath}
\usepackage{amssymb}
\usepackage{graphicx}
\usepackage{caption}
\usepackage{color}
\usepackage[colorlinks]{hyperref}
\usepackage{lscape}
\begin{document}
	\title{Detection of genuine tripartite entanglement by multiple sequential observers}

\author{Ananda G. Maity}
\email{anandamaity289@gmail.com}
\affiliation{S. N. Bose National Centre for Basic Sciences, Block JD, Sector III, Salt Lake, Kolkata 700 106, India}

\author{Debarshi Das}
\email{dasdebarshi90@gmail.com}
\affiliation{Centre for Astroparticle Physics and Space Science (CAPSS),
Bose Institute, Block EN, Sector V, Salt Lake, Kolkata 700 091, India}

\author{Arkaprabha Ghosal}
\email{a.ghosal1993@gmail.com}
\affiliation{Centre for Astroparticle Physics and Space Science (CAPSS),
Bose Institute, Block EN, Sector V, Salt Lake, Kolkata 700 091, India}

\author{Arup Roy}
\email{arup145.roy@gmail.com}
\affiliation{S. N. Bose National Centre for Basic Sciences, Block JD, Sector III, Salt Lake, Kolkata 700 106, India}

\author{A. S. Majumdar}
\email{archan@bose.res.in}
\affiliation{S. N. Bose National Centre for Basic Sciences, Block JD, Sector III, Salt Lake, Kolkata 700 106, India}

\begin{abstract}
We explore the possibility of multiple usage of a single genuine entangled state by considering a scenario consisting of three spin-$\frac{1}{2}$ particles shared between Alice, Bob and multiple Charlies. Alice performs measurements on the first particle, Bob performs measurements on the second particle and multiple Charlies perform measurements on the third particle sequentially. Here the choice of measurement settings of each Charlie is independent and uncorrelated with the choices of measurement settings and outcomes of the previous Charlies. In this scenario, we investigate whether more than one Charlie can detect genuine tripartite entanglement, and we answer this question affirmatively. In order to probe genuine entanglement, we use correlation inequalities whose violations certify genuine tripartite entanglement in a device-independent way. We extend our investigation by using appropriate genuine tripartite entanglement witness operators. Using each of these different tools for detecting genuine tripartite entanglement, we find out the maximum number of Charlies who can detect genuine entanglement in the above scenario.

\end{abstract}

\maketitle

\section{INTRODUCTION}

Entanglement \cite{ent} is one of the most fascinating non-classical features of quantum mechanics. The demarcation between separable and entangled states is well understood in bipartite scenario. But the situation becomes complex in multipartite scenarios as one can consider entanglement across many possible bipartitions. Moreover, the concept of genuine entanglement \cite{Guh} appears in the multipartite context. A multipartite state is called genuinely entangled iff it is not separable with respect to any partition. The concept of genuine entanglement is not only important for quantum foundational research, but also finds various information theoretic implications, for example, in extreme spin squeezing \cite{Sor}, high sensitive metrology tasks \cite{Hyl,toth2}, quantum computation using cluster states \cite{Rau}, measurement-based quantum computation \cite{Bri} and multiparty quantum networks \cite{Mur,Hil,Sca,Zha}. 

In spite of various successful attempts for the generation and detection of genuine multipartite entangled states \cite{Yao,Gao,Mon}, the complication  of the process is appreciated as the detection or verification of entanglement involves tomography or constructions of entanglement witnesses under precise experimental control over the system subjected to measurements. Due the difficulties present in generating genuine entanglement which is the resource for a vast range of information processing tasks, it is a significant question to ask whether genuine entanglement can be preserved partially even after performing a few cycles of local operations. The motivation of the present paper is to address the above question, and we are able to answer it in the affirmative for the tripartite scenario.

The general question as to what extent quantum correlation of an entangled state can be shared by multiple observers who perform measurements sequentially and independently of each other, was first posed in the case of the bipartite scenario.  Silva \textit{et al.} \cite{sygp}  addressed this question in the context of Bell nonlocality \cite{Bell,CHSH} by considering a scenario where an entangled pair of two spin-$\frac{1}{2}$ particles are shared between Alice in one wing and multiple Bobs in another wing. Alice acts on the first particle and multiple Bobs act on the 2nd particle sequentially, where Alice is spatially separated from the multiple Bobs. In this scenario, using a measurement model, which optimizes the trade-off between information gain and disturbance, it was conjectured \cite{sygp} that at most two Bobs can violate the Bell-CHSH (Bell-Clauser-Horne-Shimony-Holt) inequality \cite{Bell,CHSH} with a single Alice. This result is valid when the choice of measurement settings of each Bob is independent of the choices of measurement settings and outcomes of the previous Bobs and the frequencies of the inputs of each Bob are the same. This result that was subsequently confirmed analytically \cite{majumdar} applying a one-parameter positive operator valued measurement (POVM) \cite{pb1,pb2}.

Various experiments have been performed to demonstrate this phenomena \cite{exp1,exp2}. Recently, the notion of shareability of quantum nonlocality has been extended to investigate several other kinds of quantum correlations. These include sharing of EPR steering \cite{sas, shenoy}, entanglement \cite{bera,Foletto}, steerability of local quantum coherence \cite{saunak}, Bell-nonlocality with respect to quantum violations of various other Bell type inequalities \cite{das}, and preparation contextuality \cite{pcon}. These ideas have been applied in randomness generation \cite{ran}, their classical communication cost \cite{cc}, quantum  teleportation \cite{sroy}, and random access codes \cite{rac}.

Most of the previous studies have addressed the issue of sharing quantum correlations by multiple sequential observers in the bipartite scenario. Very recently, the possibility of sequential sharing of genuine tripartite nonlocality by multiple observers has been studied \cite{Saha}. Quantum entanglement is the primary ingredient for nonlocal correlations, and in the present paper we focus our attention on the sharing of genuine multipartite entanglement. In particular,  we consider the scenario where three spin-$\frac{1}{2}$ particles are spatially separated and shared between, say, Alice, Bob and multiple Charlies. Alice measures on the first particle; Bob measures on the second particle and multiple Charlies measure on the third particle sequentially. In this scenario we investigate how many Charlies can detect genuine tripartite entanglement.
 
In order to detect entanglement, one may consider the violation of Bell-type inequalities as a criterion, since  entanglement is a necessary resource for generating nonlocal correlations. One can construct inequalities which can certify genuine multipartite entanglement from the statistical data alone. This method of device-independent detection of genuine entanglement was first introduced in \cite{See,Nag,Uff,Sev} followed by an extensive formalization by Bancal \textit{et al.} \cite{Ban}. Pal \cite{Pal} and Liang \textit{et al.} \cite{Lia} have improved the existing inequalities for detecting genuine multipartite entanglement. The Mermin polynomial \cite{Mer} which is a useful tool for device-independent entanglement-witness can  be used to detect genuine tripartite entanglement  \cite{See}. In the present study, we  use quantum violations of the Mermin inequality \cite{Mer} and the Uffink inequality \cite{Uff}, respectively, in order to probe detection of genuine tripartite entanglement by multiple sequential Charlies. 

Another well developed tool for detection of entanglement is through the entanglement witness operators \cite{Horodecki,lewen,terhal2,terhal,dagmar-wit,ew}. For each entangled state, there always exists a witness operator which is a consequence of the Hahn-Banach theorem \cite{hbtheorem}. A similar concept has been formulated for the genuine tripartite entangled states (W-state and GHZ-state) which distinguishes genuine entanglement from the set of all bi-separable states \cite{Acin, Bou,Bruss,Guh}. In the present paper, we further analyse the idea of sequential detection of genuine tripartite entanglement using appropriate witness operators. 

All our analyses point out that it is indeed possible  to detect genuine entanglement sequentially by more that one Charlie. In particular, we show that at most two Charlies can detect genuine entanglement sequentially using the linear as well as nonlinear device-independent genuine entanglement inequalities. On the other hand, through appropriate genuine entanglement witnesses which are suitable for the W-state and the GHZ-state, at most four Charlies and twelve Charlies can respectively, detect genuine entanglement. Hence, this result can be useful in recycling genuine multipartite entangled resources in the context of various information processing tasks.
 
The paper is organized as follows: in Section \ref{s2} we present the basic tools for detecting genuine tripartite entanglement. The measurement scenario involving multiple sequential observers used in this paper is also described in this Section. In Section \ref{s3}, we present the main results of this paper, namely, sequential detection of genuine tripartite entanglement. Finally, we conclude in Section \ref{s4}.

\section{Preliminaries} \label{s2}
In this Section we will present some basic tools which will be used in our paper. We will also elaborate on the scenario in which sequential detection of genuine tripartite entanglement is studied.

\subsection{Detection of Genuine Entanglement}
In order to certify genuine entanglement in a device-independent way, several inequalities have been proposed. For the purpose of the present paper, we will use some of them.
A tripartite state $\rho$ is said to be bi-separable if and only if it can be written in the following form,
 \begin{equation}
\rho=\sum_{\lambda} p_{\lambda} \rho_{\lambda} ^{A} \otimes \rho_{\lambda} ^{BC} + \sum_{\mu} p_{\mu} \rho_{\mu} ^{B} \otimes \rho_{\mu} ^{AC} + \sum_{\nu} p_{\nu} \rho_{\nu} ^{C} \otimes \rho_{\nu} ^{AB},
\label{bisep}
\end{equation}
 with $0\,\leq\,p_{\lambda},\,p_{\mu},p_{\nu}\,\leq\,1$ and $\sum_{\lambda}p_{\lambda}+\sum_{\mu}p_{\mu}+\sum_{\nu}p_{\nu}=1$. A tripartite state is called genuinely entangled if and only if it cannot be written in the bi-separable form (\ref{bisep}).
 
 Let us begin with presenting the device-independent entanglement-witness provided by the Mermin polynomial \cite{Mer} as the simplest example for detecting genuine tripartite entanglement \cite{See}. Consider that three spatially separated parties, say, Alice, Bob and Charlie are sharing some quantum system in the state $\rho$. The choices of measurement settings, performed by Alice, Bob and Charlie on the shared state $\rho$ are denoted by $A_x$, $B_y$ and $C_z$ respectively, where $x$, $y$, $z$ $\in$ $\{0,1\}$. The outcomes of Alice, Bob and Charlie's measurements are denoted by $a$, $b$ and $c$, respectively, with $a$, $b$, $c$ $\in$ $\{+1, -1\}$. By repeating the experiment a number of times, the joint probability distributions $P(a, b, c|x, y, z)$ are produced. In this scenario, the Mermin inequality, whose violation certifies the presence of genuine entanglement in a device-independent way, can be expressed as \cite{See,Biswajit}:
 \begin{align}
 M =& |\langle A_1B_0C_0\rangle + \langle A_0B_1C_0\rangle + \langle A_0B_0C_1\rangle - \langle A_1B_1C_1 \rangle| \nonumber \\
 & \leq 2\sqrt{2}
 \label{m}
 \end{align}
 Here $\langle A_xB_yC_z \rangle=\sum_{abc} \, a \, b  \, c \, P(a, b, c|x, y, z)$. Here it may be noted that the violation of the inequality initially proposed by Mermin \cite{Mer} (which is nothing but $M \leq 2$) in general, does not detect genuine entanglement. Subsequently, the above inequality (\ref{m}) has been derived in order to detect genuine multipartite entanglement \cite{See,Biswajit}. Since quantum violation of the above inequality (\ref{m}) can be detected by observing the outcome statistics of the local measurements alone, it enables detecting genuine entanglement without considering the dimension of the corresponding Hilbert space, and is hence, device independent. 

With the motivation of getting stronger device-independent genuine entanglement witness, Uffink designed another nonlinear Bell-type inequality \cite{Uff} which may distinguish genuine multipartite entanglement from lesser entangled states: 
\begin{align}
U= &\langle A_1B_0C_0 +  A_0B_1C_0 + A_0B_0C_1 -  A_1B_1C_1 \rangle^2 \nonumber \\
& + \langle A_1B_1C_0 +  A_0B_1C_1 + A_1B_0C_1 -  A_0B_0C_0 \rangle^2 \leq 8.
\label{m1}
\end{align}

So far we have discussed the detection of genuine entanglement by looking at the measurement statistics in a device-independent way. However, there exist scenarios in which the devices are trusted, and one need not resort to the more resource consuming method of device-independent entanglement verification. We now describe the concept of witness operators which can also be used to detect genuine entanglement. A witness operator $\mathcal{W}$ which detects genuine entanglement of a state $\rho$ is a hermitian operator that satisfies the conditions,
 \begin{align}
 \text{Tr}(\mathcal{W}\rho)\geq 0,~~~~~ \forall \rho \in \mathcal{BS} \nonumber \\
\exists  ~~ \text{at least one }\rho \notin \mathcal{BS},~~~ \text{s.t.}~~ \text{Tr}(\mathcal{W} \rho) < 0
\end{align}  
where $\mathcal{BS}$ is the set of all bi-seperable states. The existence of such a witness operator is a consequence of the Hahn-Banach theorem on normed linear spaces \cite{hbtheorem}. For every genuinely entangled state, there exists a genuine entanglement witness.

In the present study we consider two types of witness operators that detect genuine entangled states. The first witness operator that we will use is suitable for detecting genuine entanglement of the three-qubit W-state. Consider the three-qubit W state given by, $| W \rangle = \frac{1}{\sqrt{3}}(| 001\rangle + |010\rangle + |100 \rangle)$. 
The witness operator that detects genuine entanglement in the state $|W\rangle$ is given by \cite{Acin, Bou,Bruss,Guh},
\begin{eqnarray}
&&\mathcal{W}_W = \frac{2}{3}\mathbb{I}_3 - | W \rangle \langle W |.
\label{www}
\end{eqnarray}
 Whenever a state $\rho$ gives Tr$[\mathcal{W}_W \rho] < 0$, genuine entanglement in the state $\rho$ is certified.

Next we  discuss the witness operator which is suitable for detecting genuine entanglement of three-qubit GHZ-state. Consider the three-qubit GHZ state given by, $| GHZ \rangle = \frac{1}{\sqrt{2}}(| 000\rangle + |111\rangle ) $.
The witness operator that detects genuine entanglement in the state $|GHZ\rangle$ is given by \cite{Acin,Guh}
\begin{eqnarray}
&&\mathcal{W}_{GHZ} = \frac{1}{2}\mathbb{I}_3 - | GHZ \rangle \langle GHZ |.
\label{GHZ}
\end{eqnarray} 
If a state $\rho$ gives Tr$[\mathcal{W}_{GHZ} \rho] < 0$, then genuine entanglement in the state $\rho$ is certified. 

The advantage of such kind of witness operators is that they can be implemented in the laboratory by performing a finite number of correlated local measurements. Hence, such witness operators can be realized when the observers sharing the quantum state are spatially separated. Both the  witness operators  (\ref{www},\ref{GHZ}) can be written as a sum of tensor products of local operations. The explicit forms of the decompositions of the two operators $\mathcal{W}_{W}$ and $\mathcal{W}_{GHZ}$  can be found in the Appendices \ref{appendix1} and \ref{appendix2} respectively.

\subsection{Setting up the measurement context} \label{scenario}

\begin{figure}[t!]
\centering
\includegraphics[scale=0.4]{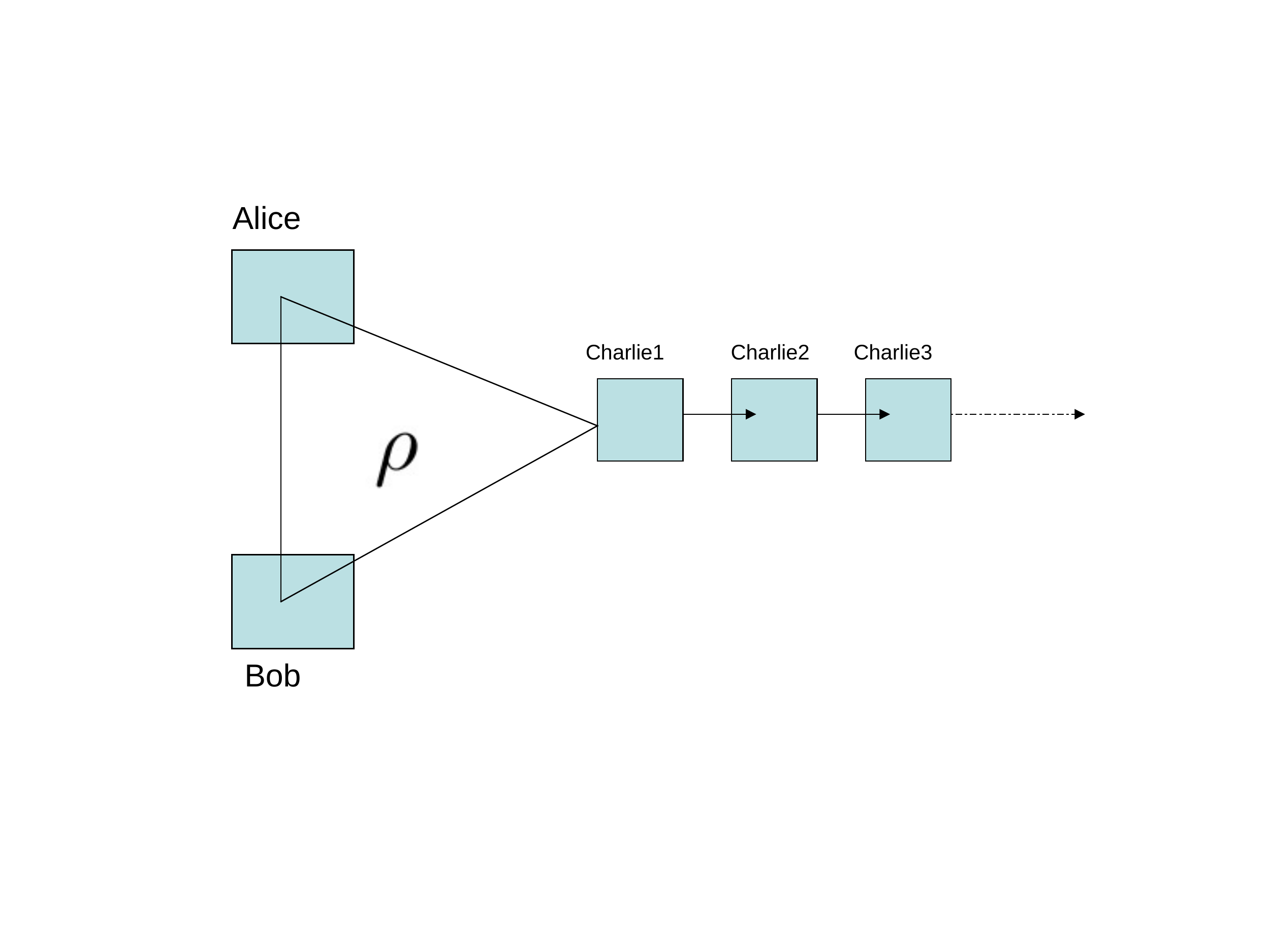}
\caption{(Color Online) Sequential detection of genuine tripartite entanglement: consider a scenario where three spin-$\frac{1}{2}$ particles are prepared in the state $\rho$. Initially $\rho$ is shared between three spatially separated parties say, Alice, Bob and Charlie$^1$. Alice as well as Bob perform projective measurements on their respective particles but Charlie$^1$ performs unsharp measurements and sends the particle to  Charlie$^2$. After doing a measurement on his respective part Charlie$^2$ again sends the particle to Charlie$^3$. In this way the protocol goes on.}
\label{fig1}
\end{figure}

In this subsection we  describe the scenario adopted in the present paper. Let us consider that three spatially separated observers say Alice, Bob and a sequence of multiple Charlies (i.e., Charlie$^1$,  Charlie$^2$, Charlie$^3$, ..., Charlie$^n$) share a  tripartite state $\rho$ consisting of three spin-$\frac{1}{2}$ particles. In our scenario, Alice performs projective measurements on the first particle, Bob performs projective measurements on the second particle and multiple Charlies are allowed to perform non-projective or unsharp measurements \cite{pb1,pb2} on the third particle sequentially. Let us now clarify the measurement scenario of multiple Charlies. Initially, Charlie$^1$  performs an unsharp measurement on the third particle, then he sends that particle to Charlie$^2$. Charlie$^2$ subsequently passes the third particle to Charlie$^3$ after performing another unsarp measurement. Charlie$^3$ also follows the same procedure and so on. This scenario is depicted in Figure \ref{fig1}.

It may be noted here that the choice of measurement settings of each Charlie is independent and uncorrelated with the choices of measurement settings and outcomes of the previous Charlies. The unbiased input scenario is another assumption that we have adopted in this paper. It implies that  all possible measurement settings of each Charlie are equally probable. Note also, that the no-signaling condition (the probability of obtaining one party's outcome does not depend on the other spatially separated party's setting) is satisfied between Alice, Bob and any Charlie as they are spatially separated and they perform measurements on three different particles. However, the no-signalling condition is not satisfied between different Charlies as each subsequent Charlie perform measurements on the same particle accessed earlier by the previous Charlie. 

In the above scenario, we ask the question as to how many Charlies can detect genuine tripartite entanglement with  Alice and  Bob. We will address this issue by investigating how many Charlies can have correlations with Alice and Bob such that they violate the Mermin inequality (\ref{m}) or the Uffink inequality (\ref{m1}). Furthermore, we will also discuss how many Charlies can demonstrate genuine tripartite entanglement if they use the witness operators given by Eq. (\ref{www})  and Eq. (\ref{GHZ}), respectively. Here, if any Charlie performs projective measurements, then the entanglement of the state will be completely lost, and there will be no chance to detect entanglement by the subsequent Charlies.  However, it is natural that no such restriction is required for the measurements performed by the last Charlie in the sequence. Hence, in order to deal with the above problem with $n$ Charlies, the first $(n - 1)$ Charlies in the sequence should perform unsharp measurements. In the following we will briefly discuss the unsharp measurement formalism used in this paper (For details, see \cite{sygp,majumdar,sas}).

Following the standard projective measurement scheme proposed by von Neuman \cite{jvn}, after an interaction with a meter having the state $\phi (q)$ ($q$ is the position of the pointer), the state $| \psi \rangle = a | 0 \rangle + b | 1 \rangle$ ($|0\rangle$ and $|1\rangle$ form orthonormal basis in $\mathbb{C}^2$, $|a|^2+|b|^2 = 1$) of the system (to be measured) of a spin-$\frac{1}{2}$ particle becomes
\begin{equation}
| \psi \rangle \otimes \phi (q) \rightarrow a | 0 \rangle \otimes \phi (q-1)  + b | 1 \rangle  \otimes \phi (q+1).
\end{equation}

In a general sharp or projective measurement, one obtains the maximum amount of information at the cost of maximum disturbance to the state of the system. On the other hand, the disturbance to the state can be reduced by performing an unsharp measurement where one obtains less amount of information. An unsharp measurement can be characterised by two real parameters: the quality factor $F$ and the precision $G$ of the measurements. The quality factor quantifies the extend to which the initial state of the system (to be measured) remains undisturbed during  the measurement process. Mathematically, the quality factor is defined as  $F(\phi(q)) = \int_{-\infty}^{\infty} \langle \phi (q+1) | \phi (q-1) \rangle dq$. Precision $G$ quantifies the information gain due to the measurement. Mathematically, it is defined as $ G(\phi(q)) = \int_{-1}^{1} \phi ^2 (q) dq$. It is  obvious that for sharp measurement $F = 0$ and $G = 1$. An optimal pointer state is the 
 one for which one obtains the greatest precision for a given quality factor.  The information-disturbance trade-off relation for an optimal pointer is given by, $F^2 + G^2 =1$ \cite{sygp}. In other words, for dichotomic measurements on a qubit system, satisfying the condition: $F^2 +G^2 = 1$ implies that the disturbance is minimized given a certain information gain.

The above formalism can be recast in terms of unsharp measurements. Unsharp measurement is one particular class of  POVMs \cite{pb1,pb2}. POVM is nothing but set of positive operators that add to identity, i. e.,  $E \equiv \{ E_i | \sum_i E_i = \mathbb{I}, 0 <E_i \leq \mathbb{I} \}$.  Here, each of the operators $E_i$ determines the probability  $\text{Tr}[\rho E_{i}]$ of obtaining the $i^{\text{th}}$ outcome (here $\rho$ is the state of the system on which the measurement is performed).

 In order to see how the  unsharp measurement formalism \cite{pb1,pb2,un1,un2,un3,un4,un5} is connected with the one-parameter class of POVMs, consider a dichotomic observable $A$ $=$ $P_{+} - P_{-}$ with outcomes $+1$ and $-1$, where $P_{+}$ ($P_{-}$) denotes the projectors associated with the outcome $+1$ ($-1$); $P_{+} + P_{-} = \mathbb{I}$ and $P_{\pm}^2 = P_{\pm}$. Given the observable $A$, one can define a dichotomic unsharp observable  $A^{\lambda} = E^{\lambda}_+ - E^{\lambda}_-$ \cite{un4,un5} associated with the sharpness parameter $\lambda \in (0, 1]$, where $E^{\lambda}_{+} + E^{\lambda}_{-} = \mathbb{I}, \, 0 < E^{\lambda}_{\pm} \leq 1$. Here the positive operators $E^{\lambda}_{\pm}$ (also known as effect operators) are given by, 
\begin{equation}
E^\lambda_{\pm} = \lambda P_{\pm} + (1-\lambda) \frac{\mathbb{I}_2}{2}.
\end{equation}
This is obtained by mixing projective measurements with white noise. The probability of getting the outcomes $+1$ and $-1$, when the above unsharp measurement is performed on the state $\rho$, are given by $\text{Tr}[\rho E^\lambda_{+}]$ and $\text{Tr}[\rho E^\lambda_{-}]$ respectively. Note that the above positive operators can also be written in the following way,
\begin{equation*}
E^{\lambda}_{\pm}= \frac{1+\lambda}{2} P_{\pm} + \frac{1-\lambda}{2} P_{\mp}.
\end{equation*}
The expectation value of $A^{\lambda}$ for a given $\rho$ is defined as \cite{un4,un5},
\begin{align}
\langle A^{\lambda} \rangle &= \text{Tr}[\rho E^\lambda_{+}] - \text{Tr}[\rho E^\lambda_{-}] \nonumber \\
&= \text{Tr}[\rho (E^\lambda_{+} -  E^\lambda_{-})] \nonumber \\
&= \lambda \langle A \rangle,
\end{align}
where $\langle A \rangle$ is the expectation value of the observable $A$ under projective measurements. The operational meaning of the expectation value $\langle A^{\lambda} \rangle$ follows from the above equation: from the probabilities ($\text{Tr}[\rho E^\lambda_{\pm}]$) of obtaining the outcomes $\pm 1$ under unsharp measurement, one can evaluate $\langle A^{\lambda} \rangle$. Note that these probabilities under unsharp measurements can be realised in experiments \cite{une1,une2,une3,une4}. 

Using the generalized von Neumann-L\"{u}ders transformation rule \cite{pb1}, the states after the measurements, when the outcomes $+1$ and $-1$ occurs, are given by, $\dfrac{\sqrt{E^\lambda_{+}} \rho \sqrt{E^\lambda_{+}}}{\text{Tr}[E^\lambda_{+} \rho]}$ and $\dfrac{\sqrt{E^\lambda_{-}} \rho \sqrt{E^\lambda_{-}}}{\text{Tr}[E^\lambda_{-} \rho]}$ respectively. In any sequential measurement scenario, we need to gain certain information while minimally disturbing the state of the system. In case of qubits, unsharp measurements are shown to be good choice for this purpose \cite{sygp}. For the von Neumann-L\"{u}ders transformation rule, it was shown \cite{majumdar} that the quality factor and the precision associated with the above unsharp measurement formalism are given by, $F = \sqrt{1-\lambda^2}$ and $G = \lambda$. Hence, the optimality condition for information gain and disturbance, $F^2 + G^2 =1$ for qubits is compatiable with the unsharp measurement formalism \cite{majumdar,sas}. In other words, the unsharp measurement formalism along with the von Neumann-L\"{u}ders transformation rule provides the largest amount of information for a given amount of disturbance created on the state due to the measurement.

In our study we will consider that each Charlie, except the final Charlie in the sequence, performs unsharp measurements.

\section{Sequential detection of genuine tripartite entanglement in the device-independent scenario}  \label{s3}

 In this section we find out the maximum number of Charlies that can independently and sequentially detect genuine entanglement in device-independent scenario. Before proceeding, we wish to mention that it was shown \cite{Saha} earlier  that at most two Charlies can simultaneously demonstrate genuine tripartite nonlocality with a single Alice and a single Bob  in the scenario described in subsection \ref{scenario}. This was demonstrated through the quantum violation of the Svetlichny inequality \cite{svet}. Since genuine entanglement is necessary for demonstrating genuine nonlocality, we can state that at least two Charlies can simultaneously demonstrate genuine tripartite entanglement in a device-independent way with a single Alice and a single Bob  through the quantum violation of the Svetlichny inequality \cite{svet}. Next, we want to find out whether the number of Charlies, who can detect genuine entanglement sequentially, can be increased using quantum violations of Mermin inequality (\ref{m}) or  Uffink inequality (\ref{m1}) in the scenario described in subsection \ref{scenario}.

We start with the Mermin inequality (\ref{m}), which is maximally violated by tripartite GHZ state \cite{GHZ}  $\rho_{GHZ} = | \psi_{GHZ} \rangle \langle \psi_{GHZ} |$, where
\begin{equation}
|\psi_{GHZ} \rangle = \frac{1}{\sqrt{2}} ( |000 \rangle + | 111 \rangle ).
\label{ghz}
\end{equation}
Suppose a tripartite GHZ state given by Eq.(\ref{ghz}) is initially shared among Alice, Bob and multiple Charlies. Alice performs dichotomic sharp measurement of spin component observable on her part in the direction $\hat{x}_0$, or $\hat{x}_1$.  Bob performs dichotomic sharp measurement of spin component observable on his particle in the direction $\hat{y}_0$ or  $\hat{y}_1$. Charlie$^m$ (where $m$ $\in \{1, 2, ..., n\}$) performs dichotomic  unsharp measurement of spin component observable in the direction $\hat{z}_0^m$ or $\hat{z}_1^m$. The outcomes of each measurement are $\pm1$.

 The projectors associated with Alice's sharp measurement of spin component observable in the direction $\hat{x}_i$ (with $i$ $\in$ $\{0,1\}$) can be written as $P_{a|\hat{x}_i} = \dfrac{\mathbb{I}_2+a \, \hat{x}_i \cdot \vec{\sigma}}{2}$ (with $a$ being the outcome of the sharp measurement and $a$ $\in$ $\{+1, -1\}$). Similarly, the projectors associated with Bob's sharp measurement of spin component observable in the direction $\hat{y}_j$ (with $j$ $\in$ $\{0,1\}$) are given by, $P_{b|\hat{y}_j} = \dfrac{\mathbb{I}_2+b \, \hat{y}_j \cdot \vec{\sigma}}{2}$ (with $b$ being the outcome of the sharp measurement and $b$ $\in$ $\{+1, -1\}$). The directions $\hat{x}_i$ and $\hat{y}_j$ can be expressed as,
\begin{equation}
\label{alicedir}
\hat{\xi}_l = \sin \theta^{\xi}_l \cos \phi^{\xi}_l \hat{X} + \sin \theta^{\xi}_l \sin \phi^{\xi}_l \hat{Y} + \cos \theta^{\xi}_l \hat{Z},
\end{equation}
where $l \in \{0, 1\}$; $0 \leq \theta^{\xi}_l  \leq \pi$; $0 \leq \phi^{\xi}_l  \leq 2 \pi$. $\hat{X}$, $\hat{Y}$, $\hat{Z}$ are three orthogonal unit vectors in Cartesian coordinates. For Alice $\xi = x$ and for Bob $\xi=y$.

The effect operators associated with Charlie$^m$'s ($m$ $\in$ $\{1, 2, ..., n \}$) unsharp measurement of spin component observable in the direction $\hat{z}^m_k$ (with $k$ $\in$ $\{0,1\}$) are given by,
\begin{equation}
E^{\lambda_m}_{c^m|\hat{z}^m_k} = \lambda_{m}\frac{\mathbb{I}_2+c^{m} \hat{z}^m_k \cdot \vec{\sigma}}{2}+(1-\lambda_{m})\frac{\mathbb{I}_2}{2},
\end{equation}
where $c^m$ is the outcome of the unsharp measurement by Charlie$^m$ and $c^m$ $\in$ $\{+1, -1\}$; $\lambda_m$ (with $0 < \lambda_m \leq 1$) denotes the sharpness parameter associated with Charlie$^m$'s unsharp measurement. When we consider a sequence of $n$ Charlies, then the measurements of Charlie$^n$ will be sharp, i.e., $\lambda_n$ = $1$. The the direction $\hat{z}^m_k$ is expressed as
\begin{equation}
\label{charliemdir}
\hat{z}^m_k = \sin \theta^{z^m}_k \cos \phi^{z^m}_k \hat{X} + \sin \theta^{z^m}_k \sin \phi^{z^m}_k \hat{Y} + \cos \theta^{z^m}_k \hat{Z},
\end{equation}
where $0 \leq \theta^{z^m}_k  \leq \pi$; $0 \leq \phi^{z^m}_k  \leq 2 \pi$.\\

The joint probability distribution of occurrence of the outcomes $a$, $b$, $c^1$, when Alice, Bob perform projective measurements of spin component observables along the directions $\hat{x}_i$ and $\hat{y}_j$ respectively, and Charlie$^1$ performs unsharp measurement of spin component observable along the direction $\hat{z}^1_k$, is given by,
\begin{align}
&P(a, b, c^1|\hat{x}_i, \hat{y}_j, \hat{z}^1_k) \nonumber \\
&=\text{Tr}\Bigg[\Bigg\{ \frac{\mathbb{I}_2+ a \hat{x}_i \cdot \vec{\sigma}}{2} \otimes \frac{\mathbb{I}_2 + b \hat{y}_j \cdot \vec{\sigma}}{2} \otimes E^{\lambda_1}_{c^1|\hat{z}^1_k} \Bigg\}  \cdot \rho_{GHZ} \Bigg].
\end{align}
The correlation function between Alice, Bob and Charlie$^1$, when Alice, Bob perform projective measurements of spin component observables along the directions $\hat{x}_i$ and $\hat{y}_j$ respectively and Charlie$^1$ performs unsharp measurement of spin component observable along the direction $\hat{z}^1_k$, can be written as 
\begin{equation}
C^1_{i,j,k}=\sum_{a = -1}^{ +1} \sum_{b =  -1}^{ +1} \sum_{c^1 = -1}^{ +1} a \, b \, c^1 \, P(a, b, c^1|\hat{x}_i, \hat{y}_j, \hat{z}^1_k).
\end{equation} 
The left hand side of the Mermin inequality (\ref{m}) associated with Alice, Bob and Charlie$^1$ in terms of the correlation functions is expressed as
\begin{equation}
M_1 = |C^1_{100}+C^1_{010}+C^1_{001}-C^1_{111}|.
\end{equation}
Now it is observed that Alice, Bob and Charlie$^1$ get quantum violation of Mermin inequality (\ref{m}) (i.e., $M_1 > 2\sqrt{2}$) when $\lambda_{1} > \frac{1}{\sqrt{2}}$. This happens for the following choice of 
measurement settings: $( \theta^x_0$, $\phi^x_0$, $\theta^x_1$, $\phi^x_1$, $\theta^y_0$, $\phi^y_0$, $\theta^y_1$, $\phi^y_1$, $\theta^{z^1}_0$, $\phi^{z^1}_0$, $\theta^{z^1}_1$, $\phi^{z^1}_1 )$ $\equiv$ $(\frac{\pi}{2}$, $\frac{\pi}{2}$, $\frac{\pi}{2}$, $0$, $\frac{\pi}{2}$, $\frac{\pi}{2}$, $\frac{\pi}{2}$, $0$, $\frac{\pi}{2}$, $\frac{\pi}{2}$, $\frac{\pi}{2}$, $0 )$.

Charlie$^1$ passes his particle to Charlie$^2$ after his measurement. The following expression gives the unnormalized post measurement reduced state at Charlie$^2$'s end after Alice, Bob get outcomes $a$, $b$ by performing projective measurements of spin component observables along the directions $\hat{x}_i$ and $\hat{y}_j$ respectively and Charlie$^1$ gets outcome $c^1$ by performing unsharp measurement of spin component observable along the direction $\hat{z}^1_k$: 
\begin{align}
\rho_{un}^{C^2} =& \text{Tr}_{A B} \Bigg[ \Bigg\{ \frac{\mathbb{I}_2+ a \hat{x}_i \cdot \vec{\sigma}}{2} \otimes \frac{\mathbb{I}_2 + b \hat{y}_j \cdot \vec{\sigma}}{2} \otimes \sqrt{E^{\lambda_1}_{c^1|\hat{z}^1_k}} \Big\}  \nonumber \\ 
& \cdot \rho_{GHZ} \cdot \Big\{ \frac{\mathbb{I}_2 + a \hat{x}_i \cdot \vec{\sigma}}{2} \otimes \frac{\mathbb{I}_2 + b \hat{y}_j \cdot \vec{\sigma}}{2} \otimes \sqrt{E^{\lambda_1}_{c^1|\hat{z}^1_k}} \Bigg\} \Bigg],
\end{align}
where,
\begin{align}
\sqrt{E^{\lambda_1}_{c^1|\hat{z}^1_k}} &= \sqrt{\dfrac{1+\lambda_1}{2}} \Bigg( \dfrac{\mathbb{I}_2 + c^1 \hat{z}^1_k \cdot \vec{\sigma}}{2} \Bigg) \nonumber \\
& +\sqrt{\dfrac{1- \lambda_1}{2}} \Bigg( \dfrac{\mathbb{I}_2 - c^1 \hat{z}^1_k \cdot \vec{\sigma}}{2} \Bigg).
\end{align}
Here $\text{Tr}_{AB}[...]$ denotes partial trace over the subsystems of Alice and Bob. On the above reduced state, Charile$^2$ again performs unsharp measurement (with sharpness parameter being denoted by $\lambda_{2}$) of spin component observable along the direction $\hat{z}^2_l$ and gets the outcome $c^2$. The joint probability distribution of occurrence of the outcomes $a$, $b$, $c^1$ $c^2$, when Alice, Bob perform projective measurements of spin component observables along the directions $\hat{x}_i$ and $\hat{y}_j$ respectively and Charlie$^1$, Charlie$^2$ perform unsharp measurement of spin component observable along the direction $\hat{z}^1_k$, $\hat{z}^2_l$ respectively, is given by,
\begin{equation}
P(a, b, c^1, c^2|\hat{x}_i, \hat{y}_j, \hat{z}^1_k, \hat{z}^2_l) = \text{Tr} \Big[ E^{\lambda_2}_{c^2|\hat{z}^2_l} \cdot \rho^{C^2}_{un} \Big].
\end{equation}
From this expression, one can obtain the joint probability of obtaining the outcomes $a$, $b$, $c^2$ when Alice, Bob, Charlie$^2$ measures spin component observables in the directions $\hat{x}_i$, $\hat{y}_j$, $\hat{z}^2_l$, respectively and when Charlie$^1$ has  already measured spin component observables in the directions $\hat{z}^1_k$,:
\begin{equation}
P(a, b, c^2|\hat{x}_i, \hat{y}_j, \hat{z}^1_k, \hat{z}^2_l) = \sum_{c^1 = -1}^{ +1} P(a, b, c^1, c^2|\hat{x}_i, \hat{y}_j, \hat{z}^1_k, \hat{z}^2_l).
\end{equation}
Let $C_{ijkl}^2$ denote the correlation between Alice, Bob and Charlie$^2$ when Alice, Bob, Charlie$^1$ and Charlie$^2$ measure spin component observables in the directions $\hat{x}_i$, $\hat{y}_j$, $\hat{z}^1_k$ and $\hat{z}^2_l$, respectively. The expression for $C_{ijkl}^2$ can be obtained from
\begin{equation}
C_{ijkl}^2 = \sum_{a = -1}^{ +1} \sum_{b =  -1}^{ +1} \sum_{c^2 = -1}^{ +1} a \, b \, c^2 \, P(a, b, c^2|\hat{x}_i, \hat{y}_j, \hat{z}^1_k, \hat{z}^2_l).
\end{equation}
Since Charlie$^2$'s choice of measurement settings is independent of the measurement settings of Charlie$^1$, the above correlation has to be averaged over the two possible measurement settings of Charlie$^1$ (spin component observables in the directions $\{ \hat{z}^1_0, \hat{z}^1_1 \}$). This average correlation function between Alice, Bob and Charlie$^2$ is given by,
\begin{equation}
\overline{C_{ijl}^2} = \sum_{k = 0,1} C_{ijkl}^2 P(\hat{z}^1_k),
\end{equation}
where $P(\hat{z}^1_k)$ is the probability with which Charlie$^1$ performs unsharp measurement of spin component observables in the direction $\hat{z}^1_k$ ($k \in \{ 0, 1 \}$). For an unbiased input scenario, we take the two measurement settings for Charlie$^1$ to be equally probable, i.e.,  $P(\hat{z}^1_0)$ = $P(\hat{z}^1_1)$ = $\frac{1}{2}$. 

The left hand side of the Mermin inequality (\ref{m}) associated with Alice, Bob and Charlie$^2$ in terms of the average correlation functions is expressed as
\begin{equation}
M_2 = |\overline{C^2_{100}}+ \overline{C^2_{010}}+ \overline{C^2_{001}}- \overline{C^2_{111}}|.
\end{equation}

In a similar way by evaluating the average correlation functions between Alice, Bob and  Charlie$^m$, the  Mermin inequality can be written as
\begin{equation}
M_m = |\overline{C_{100}^m} +  \overline{C_{010}^m} + \overline{C_{001}^m}   - \overline{C_{111}^m} | \leq 2\sqrt{2}.
\end{equation}
Violation of this inequality implies detection of genuine entanglement by Alice, Bob and Charlie$^m$.

Let us first study whether Charlie$^1$ and Charlie$^2$ can sequentially detect genuine entanglement through quantum violation of Mermin inequality (\ref{m}) with a single Alice and a single Bob in the scenario depicted in Figure \ref{fig1}. Since there are only two Charlies in this case, we consider measurements of Charlie$^{2}$ to be sharp, i.e., $\lambda_2=1$.  For the following measurement settings :  $( \theta^x_0$, $\phi^x_0$, $\theta^x_1$, $\phi^x_1$, $\theta^y_0$, $\phi^y_0$, $\theta^y_1$, $\phi^y_1$, $\theta^{z^1}_0$, $\phi^{z^1}_0$, $\theta^{z^1}_1$, $\phi^{z^1}_1$, $\theta^{z^2}_0$, $\phi^{z^2}_0$, $\theta^{z^2}_1$, $\phi^{z^2}_1 )$ $\equiv$ ($\frac{\pi}{2}$, $\frac{\pi}{2}$, $\frac{\pi}{2}$, $0$, $\frac{\pi}{2}$, $\frac{\pi}{2}$, $\frac{\pi}{2}$, $0$, $\frac{\pi}{2}$, $\frac{\pi}{2}$, $\frac{\pi}{2}$, $0$, $\frac{\pi}{2}$, $\frac{\pi}{2}$, $\frac{\pi}{2}$, $0$) and for $\lambda_1$ = $0.74$, we observe that Charlie$^1$ gets $5\%$ violation of the Mermin inequality (\ref{m}) (i.e., $M_1 = 2.96$) and Charlie$^2$ gets  $18 \%$  violation of the Mermin inequality (\ref{m}) (i.e., $M_2 = 3.34$).  Hence, Charlie$^1$ and Charlie$^2$ can detect genuine entanglement sequentially through the quantum violations of the Mermin inequality (\ref{m}). In fact, it can be shown that Charlie$^1$ and Charlie$^2$ both get quantum violations of the Mermin inequality (\ref{m}) when $\lambda_1$ $\in$ $(0.71, 0.91)$.

Next, we investigate whether Charlie$^1$, Charlie$^2$ and Charlie$^3$ can sequentially detect genuine entanglement through quantum violation of Mermin inequality (\ref{m}) with single Alice and single Bob in the scenario depicted in Figure \ref{fig1}. In this case the measurements of Charlie$^{3}$ will be sharp, i.e., $\lambda_3=1$. On the other hand, Charlie$^1$ and Charlie$^2$ perform unsharp measurements. When Charlie$^1$ gets $5\%$ violation and Charlie$^2$ gets $5\%$ violation of the Mermin inequality (\ref{m}) (i.e., when $M_1 = 2.96$ and $M_2 = 2.96$), then  the maximum magnitude of left hand side of Mermin inequality (\ref{m}) for Charlie$^3$ becomes $M_3 = 2.62$. This happens for the following choice of measurement settings:  $( \theta^x_0$, $\phi^x_0$, $\theta^x_1$, $\phi^x_1$, $\theta^y_0$, $\phi^y_0$, $\theta^y_1$, $\phi^y_1$, $\theta^{z^1}_0$, $\phi^{z^1}_0$, $\theta^{z^1}_1$, $\phi^{z^1}_1$, $\theta^{z^2}_0$, $\phi^{z^2}_0$, $\theta^{z^2}_1$, $\phi^{z^2}_1$, $\
 theta^{z^3}_0$, $\phi^{z^3}_0$, $\theta^{z^3}_1$, $\phi^{z^3}_1 )$ $\equiv$ ($\frac{\pi}{2}$, $\frac{\pi}{2}$, $\frac{\pi}{2}$, $0$, $\frac{\pi}{2}$, $\frac{\pi}{2}$, $\frac{\pi}{2}$, $0$, $\frac{\pi}{2}$, $\frac{\pi}{2}$, $\frac{\pi}{2}$, $0$, $\frac{\pi}{2}$, $\frac{\pi}{2}$, $\frac{\pi}{2}$, $0$, $\frac{\pi}{2}$, $\frac{\pi}{2}$, $\frac{\pi}{2}$, $0$)  and when $\lambda_1$ = $0.74$ and $\lambda_2$ = $0.88$. In fact, when $M_1$ = $2 \sqrt{2}$, $M_2$ = $2 \sqrt{2}$, then the maximum of $M_3$ = $2.78 < 2 \sqrt{2}$. Hence, Charlie$^1$, Charlie$^2$, Charlie$^3$ cannot detect genuine entanglement sequentially through the quantum violations of the Mermin inequality (\ref{m}). 

One important point to be noted here is that Charlie$^3$ may obtain a quantum violation of the Mermin inequality (\ref{m}) if the sharpness parameter of Charlie$^2$ or that of Charlie$^1$ is too small to get a violation. Hence, at most two Charlies can sequentially detect genuine entanglement through quantum violations of the Mermin inequality (\ref{m}).

Up to now we have used quantum violation of the Mermin inequality (\ref{m}) to certify genuine entanglement between Alice, Bob and any Charlie. Now, we will investigate whether the number of Charlies who can sequentially detect genuine entanglement, can be increased by using quantum violation of the Uffink inequality (\ref{m1}). The Uffink inequality  in terms of the average correlation functions between Alice, Bob and Charlie$^m$ can be expressed as
\begin{align}
U_m  = & \Big(\overline{C_{100}^m} + \overline{C_{010}^m} + \overline{C_{001}^m} - \overline{C_{111}^m} \Big)^2 \nonumber \\
&+ \Big(\overline{C_{110}^m} + \overline{C_{011}^m} + \overline{C_{101}^m} - \overline{C_{000}^m} \Big)^2 \leq 8.
\end{align}
The average correlation functions can be evaluated following the aforementioned procedure. Violation of this inequality implies that a genuine entangled state is shared between Alice, Bob and Charlie$^m$. In this case too, we assume that the three qubit GHZ state is initially shared between Alice, Bob and Charlie$^1$ as this state can give the maximum quantum violation ($U_m =16$) of the Uffink inequality  (\ref{m1}).

Let us try to find out whether Charlie $^1$, Charlie$^2$ and Charlie$^3$ can sequentially detect genuine entanglement through quantum violation of Uffink inequality  (\ref{m1}) with single Alice and single Bob. Here the measurements of Charlie$^{3}$ is sharp, i.e., $\lambda_3=1$.  When Charlie$^1$ gets $5\%$ violation and Charlie$^2$ gets $5\%$ violation of the Uffink inequality  (\ref{m1}) (i.e., when $U_1 = 8.40$ and $U_2 = 8.40$), then  the maximum magnitude of left hand side of Uffink inequality  (\ref{m1}) for Charlie$^3$ becomes $U_3 = 7.73$. This happens for the following choice of measurement settings: $( \theta^x_0$, $\phi^x_0$, $\theta^x_1$, $\phi^x_1$, $\theta^y_0$, $\phi^y_0$, $\theta^y_1$, $\phi^y_1$, $\theta^{z^1}_0$, $\phi^{z^1}_0$, $\theta^{z^1}_1$, $\phi^{z^1}_1$, $\theta^{z^2}_0$, $\phi^{z^2}_0$, $\theta^{z^2}_1$, $\phi^{z^2}_1$, $\theta^{z^3}_0$, $\phi^{z^3}_0$, $\theta^{z^3}_1$, $\phi^{z^3}_1 )$ $\equiv$ ($\frac{\pi}{2}$, $\frac{\pi}{2}$, $\frac{\pi}{2}$, 
 $0$, $\frac{\pi}{2}$, $\frac{\pi}{2}$, $\frac{\pi}{2}$, $0$, $\frac{\pi}{2}$, $\frac{\pi}{2}$, $\frac{\pi}{2}$, $0$, $\frac{\pi}{2}$, $\frac{\pi}{2}$, $\frac{\pi}{2}$, $0$, $\frac{\pi}{2}$, $\frac{\pi}{2}$, $\frac{\pi}{2}$, $0$)  and when $\lambda_1$ = $0.72$ and $\lambda_2$ = $0.86$. In fact, we observe that when $U_1$ = $8$, $U_2$ = $8$, then the maximum of $U_3$ = 7.76. Hence, at most two Charlies can detect genuine entanglement sequentially through the quantum violations of the Uffink inequality  (\ref{m1}).


\section{Sequential detection of genuine tripartite entanglement using witness operators} \label{s32}

In this section we are going to use genuine entanglement witnesses, instead of using device-independent genuine entanglement inequalities, in order to probe sequential detection of genuine entanglement by multiple Charlies in the scenario described in Section \ref{scenario}. The first witness operator suitable for detecting genuine entanglement of the three qubit W-state is given by \cite{Acin, Bou,Bruss,Guh},
\begin{eqnarray}
&&\mathcal{W}_W = \frac{2}{3}\mathbb{I}_3 - | W \rangle \langle W | 
\end{eqnarray}
The decomposition of this witness operator in terms of tensor products of operators is given by Eq.(\ref{decw}). However, in the scenario depicted in Figure \ref{fig1} the local measurements performed by Charlie$^m$ (except for the final Charlie in a sequence) is unsharp. Since, the decomposition (\ref{decw}) of the witness operator $\mathcal{W}_W$ can be used when each observer performs sharp projective measurements, we have to modify the decomposition (\ref{decw}) of the above witness operator for unsharp measurements at Charlie's end. In order to do this, we will follow the prescription described in \cite{bera}.

The joint probability of obtaining the outcomes $a$, $b$, $c^m$, when Alice, Bob perform projective measurements of spin component observables along the directions $\hat{x}_i$ and $\hat{y}_j$ respectively and Charlie$^m$ performs unsharp measurement of spin component observable along the direction $\hat{z}^m_k$, can be evaluated using the formula,
\begin{equation}
\mbox{Tr}\Big[\rho \Big(P_{a|\hat{x}_i}  \otimes P_{b|\hat{y}_j}  \otimes E^{\lambda_m}_{c^m|\hat{z}^m_k} \Big)\Big],
\end{equation}  
where $\rho$ is the average post-measurement state obtained after the previous stage of the measurement processes; $P_{a|\hat{x}_i}$ and $P_{b|\hat{y}_j}$ are  projection operators corresponding to the projective measurements by Alice and Bob respectively, and  $E^{\lambda_m}_{c^m|\hat{z}^m_k}$ is the effect operator associated with the POVM (unsharp measurement) performed by Charlie$^m$. 

The expectation value of the state $\rho$ corresponding to the above joint measurements is given by,
\begin{equation}
\mbox{Tr}\Big[ \Big\{ \Big(P_{+|\hat{x}_i}-P_{-|\hat{x}_i} \Big) \otimes \Big(P_{+|\hat{y}_j} - P_{-|\hat{y}_j} \Big) \otimes \Big( E^{\lambda_m}_{+|\hat{z}^m_k} - E^{\lambda_m}_{-|\hat{z}^m_k} \Big) \Big\}  \rho \Big].
\end{equation}
Now, $P_{+|\hat{x}_i}-P_{-|\hat{x}_i}$ ($P_{+|\hat{y}_j} - P_{-|\hat{y}_j}$) is nothing but $\hat{x}_i \cdot \vec{\sigma}$ ($\hat{y}_j \cdot \vec{\sigma}$). Let us denote it by $\sigma_{x_i}$ ($\sigma_{y_j}$). Let us also denote $E^{\lambda_m}_{+|\hat{z}^m_k} - E^{\lambda_m}_{-|\hat{z}^m_k}$ as $\sigma^{\lambda_m}_{z^m_k}$. Hence, we can write the following,
\begin{widetext}
\begin{eqnarray}
&&\langle \sigma_{x_i} \otimes \sigma_{y_j} \otimes \sigma^{\lambda_m}_{z^m_k} \rangle \nonumber \\
&=& \mbox{Tr}\Big[ \Big(P_{+|\hat{x}_i}-P_{-|\hat{x}_i} \Big) \otimes \Big(P_{+|\hat{y}_j} - P_{-|\hat{y}_j} \Big)  \otimes \Big( E^{\lambda_m}_{+|\hat{z}^m_k} - E^{\lambda_m}_{-|\hat{z}^m_k} \Big)  \rho \Big] \nonumber\\
&=&\mbox{Tr}\Big[ \Big(P_{+|\hat{x}_i}-P_{-|\hat{x}_i} \Big) \otimes \Big(P_{+|\hat{y}_j} - P_{-|\hat{y}_j} \Big)  \otimes \lambda_m \Big( P_{+|\hat{z}^m_k}-P_{-|\hat{z}^m_k} \Big)  \rho \Big] \nonumber\\
&=&\lambda_m \langle \sigma_{x_i} \otimes \sigma_{y_j} \otimes \sigma_{z^m_k} \rangle.
\end{eqnarray}
\end{widetext}
Noting the above relation,  one can use the substitution $\langle \sigma_{x_i} \otimes \sigma_{y_j} \otimes \sigma^{\lambda_m}_{z^m_k} \rangle  \rightarrow \lambda_m \langle \sigma_{x_i} \otimes \sigma_{y_j} \otimes \sigma_{z^m_k} \rangle$ in the case of a general $\lambda_m$ \cite{bera}.  The unsharp version decomposition (\ref{decw}) of the genuine entanglement witness operator $\mathcal{W}^{\lambda_m}_W$ using the above substitution is given in the Appendix \ref{appendix3}.

Now, since we have Tr$[\mathcal{W}_W \rho_{BS}] \geq 0$ $\forall$ $\rho_{BS}$ $\in$ $ \mathcal{BS}$ (where $\mathcal{BS}$ is the set of all bi-seperable states) and $0 < \lambda_m \leq 1$, we have $\mbox{Tr}[\mathcal{W}_W^{\lambda_m} \rho_{BS}] \geq 0$ $\forall$ $\rho_{BS}$ $\in$ $ \mathcal{BS}$ (The detailed proof is given in the Appendix \ref{appendix3}).  Hence, we can conclude that the operator $\mathcal{W}_W$ even after introducing unsharpness in Charlie's measurements ($\mathcal{W}_W^{\lambda_m}$) can be used as a valid witness of genuine entanglement. 

{\centering
\begin{table}[ht]
 \begin{tabular}{||c | c ||} 
 \hline 
 Charlie$^m$ &Permissible ranges for $\lambda_m$ \\ [0.5ex] 
 \hline\hline
 Charlie$^1$ & $ 1 \geq \lambda_1 > \lambda_1^{min} = 0.54 $  \\ 
 \hline
 Charlie$^2$ & 1 $\geq \lambda_2 > \lambda_2^{min} = 0.60 $  \\
 & when $\lambda_i = \lambda_i^{min}$ $\forall$ $i < 2$ \\
 \hline
 Charlie$^3$ & $ 1 \geq \lambda_3 > \lambda_3^{min} = 0.69 $  \\
 & when $\lambda_i = \lambda_i^{min}$ $\forall$ $i < 3$  \\
 \hline
 Charlie$^4$ & $ 1 \geq \lambda_4 > \lambda_4^{min} = 0.84 $  \\
 & when $\lambda_i = \lambda_i^{min}$ $\forall$ $i < 4$ \\
 \hline
 Charlie$^5$ & No valid permissible range for $\lambda_5$ \\
 & (since $0 < \lambda_5 \leq 1$) \\
 & when $\lambda_i = \lambda_i^{min}$ $\forall$ $i < 5$ \\ [1ex] 
 \hline
\end{tabular}
\caption{Here we show the permissible ranges of  sharpness parameters $\lambda_m$ (where $0 < \lambda_m \leq 1$)  of Charlie$^m$  for detecting genuine entanglement through the witness operator $\mathcal{W}_W^{\lambda_m}$ with a single Alice and a single Bob at the other sides. The permissible range of each $\lambda_m$ depends on the values  $\lambda_1$, $\lambda_2$, ..., $\lambda_{m-1}$. In the above table we have presented the permissible range of each $\lambda_m$ for the minimum permissible values of $\lambda_1$, $\lambda_2$, ..., $\lambda_{m-1}$. For other values of $\lambda_1$, $\lambda_2$, ..., $\lambda_{m-1}$, the permissible range of each $\lambda_m$ can also be calculated. However, the permissible ranges of $\lambda_m$ will be smaller than that presented in the table if we take other values $\lambda_i > \lambda_i^{min}$ $\forall$ $i < m$, and the maximum number of Charlies may get reduced. Here we find that at most four Charlies can detect genuine tripartite entanglement through the witness operator $\mathcal{W}_W^{\lambda_m}$.}
\label{tab1}
\end{table}
}

 Before proceeding further, let us elaborate on the scenario used by us in the context of genuine entanglement witness operators. As discussed earlier, we have considered that Alice, Bob and any Charlie are spatially separated from each other. The no-signaling condition is satisfied between Alice, Bob and any Charlie.  Each of Alice, Bob and Charlie$^m$ (where $m \geq 1$ is arbitrary) always performs any of the pre-defined (sharp or unsharp) measurements (e.g. in case of $\mathcal{W}^{\lambda_m}_{W}$, $\sigma_z$, $(\sigma_z + \sigma_x)/\sqrt{2}$, $(\sigma_z - \sigma_x)/\sqrt{2}$, $(\sigma_z + \sigma_y)/\sqrt{2}$, 
 $(\sigma_z - \sigma_y)/\sqrt{2}$) randomly in any experimental run.  After completion of the experiment, they communicate their choice of measurement setting and outcome for each of the experimental run to the referee. The referee then determines the correlations necessary to evaluate the witness. For example, when the referee wants to determine $\langle \sigma_z\otimes \sigma_z \otimes \lambda_m \, \sigma_z \rangle$ between Alice, Bob and Charlie$^m$, then the referee will only consider the data of those experimental runs in which each of Alice, Bob and Charlie$^m$ performs (sharp or unsharp) measurement of $\sigma_z$.
 
In case of $\mathcal{W}^{\lambda_m}_{W}$, Charlie$^m$ always measures unsharp version of any of the above-mentioned five observables. But the choice of the particular measurement (among the above five measurements) performed by Charlie$^m$ in a run is always independent of the particular measurements performed by previous Charlies. To summarize, any Charlie always performs one of the above-mentioned five measurements randomly and independently of the choices of measurements (among the above-mentioned five measurements) of the previous Charlies.

Now, for example, consider an experimental run in which Charlie$^1$ and Charlie$^2$ perform unsharp measurements of $\sigma_z$ and $(\sigma_z + \sigma_x)/\sqrt{2}$ respectively. Then Charlie$^1$'s measurement outcome will be useful for the referee to calculate $\langle \sigma_z\otimes \sigma_z \otimes \lambda_1 \, \sigma_z \rangle$ between Alice, Bob and Charlie$^1$ if Alice and Bob perform sharp measurements of $\sigma_z$ in that experimental run. But Charlie$^2$'s measurement outcomes will not be useful in this case to calculate any of the correlations necessary to evaluate the witness. On the other hand, Charlie$^2$'s measurement outcome in the above experimental run will be useful for the referee to calculate $\langle (\sigma_z + \sigma_x) \otimes (\sigma_z + \sigma_x) \otimes \lambda_2 \, (\sigma_z + \sigma_x) \rangle$ if Alice and Bob perform sharp measurements of $(\sigma_z + \sigma_x)/\sqrt{2}$ in that experimental run. But Charlie$^1$'s measurement outcomes will not be useful in this case. 

Now, suppose that the three qubit W state given by, $| W \rangle = \frac{1}{\sqrt{3}}(| 001\rangle + |010\rangle + |100 \rangle)$ is initially shared between Alice, Bob and Charlie$^1$. When Alice, Bob perform projective measurements and Charlie$^1$ performs unsharp measurement with sharpness parameter being denoted by $\lambda_1$, the entanglement witness $\mathcal{W}_{W}^{\lambda_1}$ acquires the following expectation value
\begin{equation}
\mbox{Tr}\Big[ | W \rangle \langle W |~ \mathcal{W}_W^{\lambda_1}\Big] = \frac{1}{18}(7-13\lambda_1)
\label{c1witw}
\end{equation}
It is clear from the above equation that Charlie$^1$ can detect genuine entanglement with Alice and Bob when $\lambda_1 > \frac{7}{13} \simeq 0.54$.

Let us now explore whether there is any possibility for subsequent Charlies, i.e, Charlie$^2$, Charlie$^3$ ...., to detect the residual genuine entanglement in the post measurement average state with single Alice and single Bob at other sides. Since any Charlie is ignorant about the choices of measurement settings and outcomes all previous Charlies, we have to average over the previous Charlie's inputs and outputs to obtain the state shared between Alice, Bob and the Charlie of the current stage  of the experiment. After performance of Charlie$^1$'s unsharp measurement, the average state becomes, 
\begin{align}
&|W \rangle \langle W | \to \rho_W^{\lambda_1}  \nonumber \\
&=\frac{1}{5} \sum_{i, \hat{z}^1_k} \Big( \mathbb{I} \otimes \mathbb{I} \otimes \sqrt{E^{\lambda_1}_{i|\hat{z}^1_k}} \Big) |W \rangle \langle W |  \big(\mathbb{I} \otimes \mathbb{I} \otimes \sqrt{E^{\lambda_1}_{i|\hat{z}^1_k}} \Big),
\end{align}
where $i$ $\in$ $\{+1, -1\}$, $\hat{z}^1_k$ $\in$ $\Big\{\hat{z},\dfrac{\hat{z}+\hat{x}}{\sqrt{2}}$, $\dfrac{\hat{z}-\hat{x}}{\sqrt{2}}$, $\dfrac{\hat{z}+\hat{y}}{\sqrt{2}}$, $\dfrac{\hat{z}-\hat{y}}{\sqrt{2}}\Big\}$.

In the next step Charlie$^2$ performs unsharp measurements on his part of $\rho_W^{\lambda_1}$ with sharpness parameter $\lambda_2$, to check with Alice and Bob whether the state is genuinely entangled, by using the witness parameter $\mathcal{W}_W^{\lambda_2}$ which acquires the following expectation value, 
\begin{align}
\mbox{Tr}\Big[ \rho_W^{\lambda_1} \mathcal{W}_{W}^{\lambda_2}\Big] = \frac{1}{90}\Big(35-\big(23+42\sqrt{1-\lambda_1^2} \big)\lambda_2 \Big).
\end{align} 
Hence, Charlie$^2$ can detect genuine entanglement with Alice and Bob if $\dfrac{1}{90}\Big(35-\big(23+42\sqrt{1-\lambda_1^2} \big)\lambda_2 \Big)$ $< 0$. On the other hand, from Eq.(\ref{c1witw}) we know that Charlie$^1$ can detect genuine entanglement with Alice and Bob when $\lambda_1 > \frac{7}{13}$, i.e.,  when $\lambda_1=\frac{7}{13}+\epsilon$ with $\epsilon$ being a positive number such that $\epsilon \leq \frac{6}{13}$. Hence, in order to detect genuine entanglement, Charlie$^2$ must choose his sharpness parameter $\lambda_2$ such that it satisfies $\frac{1}{90}\Big(35-\big(23+42\sqrt{1-(\frac{7}{13}+\epsilon)^2} \big)\lambda_2 \Big)$ $<0$. If we take $\epsilon = 0$ (i.e., $\lambda_1=\frac{7}{13}$), then we obtain that Charlie$^2$ can detect genuine entanglement with Alice and Bob if $\lambda_2 > 0.60$.

In this way if we proceed it can be observed  that at most four Charlies can detect genuine entanglement through the witness operator $\mathcal{W}_W^{\lambda_m}$ when the initial shared state is three qubit pure W-state.  Allowed ranges of the sharpness parameters associated with different Charlies' measurements in order to detect genuine entanglement using the witness operator $\mathcal{W}_W^{\lambda_m}$ are presented in Table \ref{tab1}.

Now we are going to investigate the maximum number of Charlies that can detect genuine entanglement in the scenario mentioned in Figure \ref{fig1} using another type of witness operator (suitable for detecting genuine entanglement of three qubit GHZ state) which is given by \cite{Acin,Guh},
\begin{align}
\mathcal{W}_{GHZ} &= \frac{1}{2}\mathbb{I}_3 - | GHZ \rangle \langle GHZ | 
\end{align}
Now, when any Charlie$^m$ performs unsharp measurements with sharpness parameter $\lambda_m$, the decomposition of the genuine entanglement witness operator $\mathcal{W}^{\lambda_m}_{GHZ}$ is presented in the Appendix \ref{appendix4}.

Now, since we have Tr$[\mathcal{W}_{GHZ} \rho_{BS}] \geq 0$ $\forall$ $\rho_{BS}$ $\in$ $ \mathcal{BS}$ and $0 < \lambda_m \leq 1$, we can write that Tr$[\mathcal{W}_{GHZ}^{\lambda_m} \rho_{BS}] \geq 0$ $\forall$ $\rho_{BS}$ $\in$ $ \mathcal{BS}$ (details can be found in Appendix \ref{appendix4}). Hence, one may conclude that the operator $\mathcal{W}_{GHZ}^{\lambda_m}$ after introducing unsharpness in Charlie's measurements can again be used as a valid witness operator of genuine entanglement.

{\centering
	\begin{table}[ht]
 \begin{tabular}{||c | c  ||} 
 \hline 
 Charlie$^m$ &Permissible ranges for $\lambda_m$ \\ [0.5ex] 
 \hline\hline
 Charlie$^1$ & $ 1 \geq \lambda_1 > \lambda_1^{min} = 0.33 $ \\ 
 \hline
 Charlie$^2$ & 1 $\geq \lambda_2 > \lambda_2^{min} = 0.35 $  \\
 & when $\lambda_i = \lambda_i^{min}$ $\forall$ $i < 2$ \\
 \hline
 Charlie$^3$ & $ 1 \geq \lambda_3 > \lambda_3^{min} = 0.36 $  \\
 & when $\lambda_i = \lambda_i^{min}$ $\forall$ $i < 3$  \\
 \hline
 Charlie$^4$ & $ 1 \geq \lambda_4 > \lambda_4^{min} = 0.38 $  \\
 & when $\lambda_i = \lambda_i^{min}$ $\forall$ $i < 4$ \\
 \hline
 Charlie$^5$ & $ 1 \geq \lambda_5 > \lambda_5^{min} = 0.40 $  \\
 & when $\lambda_i = \lambda_i^{min}$ $\forall$ $i < 5$ \\
 \hline
 Charlie$^6$ & $ 1 \geq \lambda_6 > \lambda_6^{min} = 0.42 $  \\
 & when $\lambda_i = \lambda_i^{min}$ $\forall$ $i < 6$ \\
 \hline
 Charlie$^7$ & $ 1 \geq \lambda_7 > \lambda_7^{min} = 0.45 $  \\
 & when $\lambda_i = \lambda_i^{min}$ $\forall$ $i < 7$ \\
 \hline
 Charlie$^8$ & $ 1 \geq \lambda_8 > \lambda_8^{min} = 0.48 $  \\
 & when $\lambda_i = \lambda_i^{min}$ $\forall$ $i < 8$ \\
 \hline
 Charlie$^9$ & $ 1 \geq \lambda_9 > \lambda_9^{min} = 0.53 $  \\
 & when $\lambda_i = \lambda_i^{min}$ $\forall$ $i < 9$ \\
 \hline
 Charlie$^{10}$ & $ 1 \geq \lambda_{10} > \lambda_{10}^{min} = 0.59 $  \\
 & when $\lambda_i = \lambda_i^{min}$ $\forall$ $i < 10$ \\
 \hline
 Charlie$^{11}$ & $ 1 \geq \lambda_{11} > \lambda_{11}^{min} = 0.67 $  \\
 & when $\lambda_i = \lambda_i^{min}$ $\forall$ $i < 11$ \\
 \hline
 Charlie$^{12}$ & $ 1 \geq \lambda_{12} > \lambda_{12}^{min} = 0.81 $  \\
 & when $\lambda_i = \lambda_i^{min}$ $\forall$ $i < 12$ \\
 \hline
 Charlie$^{13}$ & No valid permissible range for $\lambda_{13}$ \\
 & (since $0 < \lambda_5 \leq 1$) \\
 & when $\lambda_i = \lambda_i^{min}$ $\forall$ $i < 13$ \\[1ex] 
 \hline
\end{tabular}
\caption{Here we show the permissible ranges of  sharpness parameters $\lambda_m$ (where $0 < \lambda_m \leq 1$)  of Charlie$^m$  for detecting genuine entanglement through the witness operator $\mathcal{W}_{GHZ}^{\lambda_m}$ with a single Alice and a single Bob at the other sides. The permissible range of each $\lambda_m$ depends on the values  $\lambda_1$, $\lambda_2$, ..., $\lambda_{m-1}$. In the above table we have presented the permissible range of each $\lambda_m$ for the minimum permissible values of $\lambda_1$, $\lambda_2$, ..., $\lambda_{m-1}$. For other values of $\lambda_1$, $\lambda_2$, ..., $\lambda_{m-1}$, the permissible range of each $\lambda_m$ can also be calculated. However, the permissible ranges of $\lambda_m$ will be smaller than that presented in the table if we take other values $\lambda_i > \lambda_i^{min}$ $\forall$ $i < m$, and the maximum number of Charlies may get reduced. Here we find that at most twelve Charlies can detect genuine tripartite entanglement through the witness operator $\mathcal{W}_W^{\lambda_m}$.}
\label{tab2}
\end{table}
}

In this case, consider that the three qubit GHZ state given by, $| GHZ \rangle = \frac{1}{\sqrt{2}}(| 000\rangle + |111\rangle ) $, is initially shared between Alice, Bob and Charlie$^1$. In a similar fashion described earlier, we now investigate how many Charlies can detect genuine entanglement sequentially with single Alice and single Bob. Since Alice, Bob perform projective measurements and Charlie$^1$ performs unsharp measurement with sharpness parameter $\lambda_1$, the expectation value of the genuine entanglement witness $\mathcal{W}_{GHZ}^{\lambda_1}$ becomes,
\begin{equation}
\mbox{Tr}\Big[ | GHZ \rangle \langle GHZ |~ \mathcal{W}_{GHZ}^{\lambda_1}\Big] = \frac{1}{4}(1 - 3 \lambda_1)
\end{equation}
Hence, it is clear from the above expectation value that Charlie$^1$ can detect genuine entanglement using the genuine entanglement witness $\mathcal{W}_{GHZ}^{\lambda_1}$  with Alice and Bob when $\lambda_1 > \frac{1}{3} \equiv 0.33$.

After Charlie$^1$'s unsharp measurement, the average state shared between Alice, Bob and Charlie$^2$ becomes 
\begin{align}
&|GHZ \rangle \langle GHZ | \to \rho_{GHZ}^{\lambda_1}  \nonumber \\
&=\frac{1}{4} \sum_{i, \hat{z}^1_k} \Big( \mathbb{I} \otimes \mathbb{I} \otimes \sqrt{E^{\lambda_1}_{i|\hat{z}^1_k}} \Big) |GHZ \rangle \langle GHZ |  \big(\mathbb{I} \otimes \mathbb{I} \otimes \sqrt{E^{\lambda_1}_{i|\hat{z}^1_k}} \Big),
\end{align}
where $i$ $\in$ $\{+1, -1\}$, $\hat{z}^1_k$ $\in$ $\Big\{\hat{z}$, $\hat{x}$, $\dfrac{\hat{x}+\hat{y}}{\sqrt{2}}$, $\dfrac{\hat{x}-\hat{y}}{\sqrt{2}} \Big\}$.

Next, Charlie$^2$  performs unsharp measurements on his part of $\rho_{GHZ}^{\lambda_1}$ with sharpness parameter $\lambda_2$, to check with Alice and Bob whether the state is genuinely entangled. In this case, the expectation value of the witness operator $\mathcal{W}_{GHZ}^{\lambda_2}$ becomes, 
\begin{equation}
\mbox{Tr}\Big[ \rho_{GHZ}^{\lambda_1} \mathcal{W}_{GHZ}^{\lambda_2}\Big] = \frac{1}{4} \Big[1 - \Big(1+2\sqrt{1-\lambda_1^2} \Big)\lambda_2 \Big].
\end{equation}
Hence, Charlie$^2$ can detect genuine entanglement with Alice and Bob using the above witness if $\frac{1}{4} \Big[1 - \Big(1+2\sqrt{1-\lambda_1^2} \Big)\lambda_2 \Big]$ $< 0$. Since, Charlie$^1$ can detect genuine entanglement with Alice and Bob when $\lambda_1=\frac{1}{3}+\epsilon$ with $\epsilon$ being a positive number such that $\epsilon \leq \frac{2}{3}$. Hence, for detecting genuine entanglement, Charlie$^2$ must choose his sharpness parameter $\lambda_2$ such that $\frac{1}{4} \Big[1 - \Big(1+2\sqrt{1- (\frac{1}{3}+\epsilon)^2} \Big)\lambda_2 \Big]$ $<0$. For example, if we take $\epsilon = 0$ (i.e., $\lambda_1=\frac{1}{3}$), then Charlie$^2$ can detect genuine entanglement with Alice and Bob if $\lambda_2 > 0.35$.

Next, we continue exploring the possibility for subsequent Charlies (Charlie$^3$, Charlie$^4$, ...) to detect genuine entanglement. We observe that at most twelve Charlies can detect genuine entanglement through the witness operator $\mathcal{W}_{GHZ}^{\lambda_m}$ when the initial shared state is three qubit pure GHZ  state. Allowed ranges of the sharpness parameters associated with different Charlies' measurements in order to detect genuine entanglement using the witness operator $\mathcal{W}_{GHZ}^{\lambda_m}$ are presented in Table \ref{tab2}.


\section{CONCLUSIONs} \label{s4}
There exist several communication and computational tasks where multipartite quantum correlations serve as resources \cite{Sor,Hyl,toth2,Sca,ap1,ap2,ap3,ap5,ap6,ap7,ap8,ap9,ap10,ap11}. However, due to the difficulties present in experimentally producing multipartite quantum correlations, their implementation as powerful resources in various information processing tasks are still elusive. Hence, exploring the possibilities of using single multipartite quantum correlation several times is not only interesting for  foundational studies but may also be useful for  information theoretic applications. 

In the present study we address the question as to whether multiple observers can detect genuine tripartite entanglement sequentially.  We consider the scenario where three spin-$\frac{1}{2}$ particles are spatially separated and shared between, say, Alice, Bob and multiple Charlies. Alice measures on the first particle; Bob measures on the second particle and multiple Charlies measure on the third particle sequentially.  In the course of our study we have used both linear as well as non-linear correlation inequalities which detect genuine entanglement in the device-independent scenario. In this context, we have shown that at most two Charlies can detect genuine entanglement of the GHZ-state. Note that the question of sharing of genuine entanglement of the W-state in the device-independent scenario remains to be investigated due to the lack of a suitable inequality. A possible direction in this context  may be to evaluate the bi-separable bounds of the inequalities presented in \cite{newww}, which are violated maximally by the W state.

The number of Charlies may be increased by giving up the requirement of device-independence, as we have shown using two types of appropriate genuine entanglement witness operators. Here, we find that at most four Charlies  can detect genuine entanglement sequentially with the single Alice and single Bob using the shared W-state. In case of the shared GHZ-state we find that the number of Charlies can increase up to twelve, which may open up interesting possibilities of detection of genuine tripartite entanglement sharing by multiple observers.

Before concluding, it may be noted that the issue of sharing genuine nonlocality in the above scenario has been studied earlier \cite{Saha}. Hence, it would be interesting to investigate this issue in the intermediate context between entanglement and Bell-nonlocality, {\it viz.}, sharing of genuine multipartite quantum steering \cite{stm1,stm2,stm3} by multiple observers measuring sequentially on the same particle. Finally, exploring information theoretic applications of the present study is another direction for future research.

\section{ ACKNOWLEDGEMENTS}
DD acknowledges the financial support from University Grants Commission (UGC), Government of India. DD acknowledges his visit
at S. N. Bose National Centre for Basic Sciences, Kolkata where a part of the work
was done. ASM acknowledges support from the DST project DST/ICPS/QuEST/2019/Q98.

\newpage

\begin{widetext}
\appendix

\section{Explicit decomposition of the witness operator $\mathcal{W}_W$ into a sum of tensor products of local operators}\label{appendix1}
The witness operator $\mathcal{W}_W$ given by Eq.(\ref{www}) can be decomposed as follows:
\begin{align}
&\mathcal{W}_W =  \frac{1}{24} \Big(13 \, \mathbb{I} \otimes \mathbb{I} \otimes \mathbb{I} + 3 \, \sigma_z \otimes \mathbb{I} \otimes \mathbb{I} + 3 \, \mathbb{I}\otimes \sigma_z \otimes \mathbb{I} + 3 \, \mathbb{I} \otimes \mathbb{I}\otimes \sigma_z + 5 \, \sigma_z \otimes \sigma_z \otimes \mathbb{I} + 5 \, \sigma_z \otimes \mathbb{I}\otimes \sigma_z + 5 \, \mathbb{I} \otimes \sigma_z \otimes \sigma_z \nonumber \\
&  + 7 \, \sigma_z\otimes \sigma_z \otimes \sigma_z - \mathbb{I} \otimes \mathbb{I} \otimes (\sigma_z + \sigma_x)  -  \mathbb{I}  \otimes (\sigma_z + \sigma_x) \otimes \mathbb{I} - (\sigma_z + \sigma_x) \otimes \mathbb{I} \otimes \mathbb{I} - \mathbb{I} \otimes (\sigma_z + \sigma_x) \otimes (\sigma_z + \sigma_x) \nonumber \\
&   - (\sigma_z + \sigma_x) \otimes \mathbb{I} \otimes (\sigma_z + \sigma_x) - (\sigma_z + \sigma_x)  \otimes (\sigma_z + \sigma_x) \otimes \mathbb{I}  - (\sigma_z + \sigma_x) \otimes(\sigma_z + \sigma_x) \otimes(\sigma_z + \sigma_x) - \mathbb{I} \otimes \mathbb{I} \otimes (\sigma_z - \sigma_x) \nonumber \\
& -  \mathbb{I}  \otimes (\sigma_z - \sigma_x) \otimes \mathbb{I} - (\sigma_z - \sigma_x) \otimes \mathbb{I} \otimes \mathbb{I} - \mathbb{I} \otimes (\sigma_z - \sigma_x) \otimes (\sigma_z - \sigma_x)  - (\sigma_z - \sigma_x) \otimes \mathbb{I} \otimes (\sigma_z - \sigma_x) \nonumber \\
&- (\sigma_z - \sigma_x)  \otimes (\sigma_z - \sigma_x) \otimes \mathbb{I} - (\sigma_z - \sigma_x) \otimes(\sigma_z - \sigma_x) \otimes(\sigma_z - \sigma_x)  - \mathbb{I} \otimes \mathbb{I} \otimes (\sigma_z + \sigma_y)  -  \mathbb{I}  \otimes (\sigma_z + \sigma_y) \otimes \mathbb{I} \nonumber \\
& - (\sigma_z + \sigma_y) \otimes \mathbb{I} \otimes \mathbb{I} - \mathbb{I} \otimes (\sigma_z + \sigma_y) \otimes (\sigma_z + \sigma_y)  - (\sigma_z + \sigma_y) \otimes \mathbb{I} \otimes (\sigma_z + \sigma_y) - (\sigma_z + \sigma_y)  \otimes (\sigma_z + \sigma_y) \otimes \mathbb{I} \nonumber \\
& -(\sigma_z + \sigma_y) \otimes(\sigma_z + \sigma_y) \otimes(\sigma_z + \sigma_y)  - \mathbb{I} \otimes \mathbb{I} \otimes (\sigma_z - \sigma_y)  -  \mathbb{I}  \otimes (\sigma_z - \sigma_y) \otimes \mathbb{I} - (\sigma_z - \sigma_y) \otimes \mathbb{I} \otimes \mathbb{I}  \nonumber \\
&- \mathbb{I} \otimes (\sigma_z - \sigma_y) \otimes (\sigma_z - \sigma_y)  -(\sigma_z - \sigma_y) \otimes \mathbb{I} \otimes (\sigma_z - \sigma_y) - (\sigma_z - \sigma_y)  \otimes (\sigma_z - \sigma_y) \otimes \mathbb{I}\nonumber \\
& -(\sigma_z - \sigma_y)  \otimes (\sigma_z - \sigma_y) \otimes (\sigma_z - \sigma_y) \Big).
\label{decw}
\end{align}
Note that all the correlations of measurements like $\sigma_z \otimes \sigma_z \otimes \sigma_z$,  $\sigma_z \otimes \sigma_z \otimes \mathbb{I}$,  $\sigma_z \otimes \mathbb{I} \otimes \sigma_z$,  $\mathbb{I} \otimes \sigma_z \otimes \sigma_z$,  $\sigma_z \otimes \mathbb{I} \otimes \mathbb{I}$,  $\mathbb{I} \otimes \sigma_z \otimes \mathbb{I}$,  $\mathbb{I} \otimes \mathbb{I} \otimes \sigma_z$ can be determined from the same data. Hence, the above decomposition requires measurements of five correlations: 

$\bullet$ $\sigma_z \otimes \sigma_z \otimes \sigma_z$, 

$\bullet$ $\Big(\dfrac{\sigma_z + \sigma_x}{\sqrt{2}} \Big)  \otimes \Big(\dfrac{\sigma_z + \sigma_x}{\sqrt{2}} \Big) \otimes \Big(\dfrac{\sigma_z + \sigma_x}{\sqrt{2}} \Big)$, 

$\bullet$ $\Big(\dfrac{\sigma_z - \sigma_x}{\sqrt{2}} \Big)  \otimes \Big(\dfrac{\sigma_z - \sigma_x}{\sqrt{2}} \Big)  \otimes \Big(\dfrac{\sigma_z - \sigma_x}{\sqrt{2}} \Big)$, 

$\bullet$ $\Big(\dfrac{\sigma_z + \sigma_y}{\sqrt{2}} \Big)  \otimes \Big(\dfrac{\sigma_z + \sigma_y}{\sqrt{2}} \Big)  \otimes \Big( \dfrac{\sigma_z + \sigma_y}{\sqrt{2}}\Big)$, 

$\bullet$ $\Big(\dfrac{\sigma_z - \sigma_y}{\sqrt{2}} \Big)  \otimes \Big(\dfrac{\sigma_z - \sigma_y}{\sqrt{2}} \Big) \otimes \Big(\dfrac{\sigma_z - \sigma_y}{\sqrt{2}} \Big)$. 

\section{Explicit decomposition of the witness operator $\mathcal{W}_{GHZ}$ into a sum of tensor products of local operators}\label{appendix2}
The witness operator $\mathcal{W}_{GHZ}$ given by Eq.(\ref{GHZ}) can be written in the following decomposition:
\begin{align}
\mathcal{W}_{GHZ} &= \frac{1}{8} \Big(3 \, \mathbb{I} \otimes \mathbb{I} \otimes \mathbb{I} - \mathbb{I} \otimes \sigma_z \otimes \sigma_z - \sigma_z \otimes \mathbb{I} \otimes \sigma_z -  \sigma_z \otimes \sigma_z \otimes \mathbb{I} - 2 \, \sigma_x \otimes \sigma_x \otimes \sigma_x \nonumber \\
& +  \frac{1}{2}(\sigma_x + \sigma_y) \otimes (\sigma_x + \sigma_y) \otimes (\sigma_x + \sigma_y) + \frac{1}{2}(\sigma_x - \sigma_y) \otimes (\sigma_x - \sigma_y) \otimes (\sigma_x - \sigma_y) \Big).
\label{decghz}
\end{align}
The above decomposition requires measurements of four correlations: 

$\bullet$ $\sigma_z \otimes \sigma_z \otimes \sigma_z$, 

$\bullet$ $\sigma_x \otimes \sigma_x \otimes \sigma_x$, 

$\bullet$ $\Big(\dfrac{\sigma_x + \sigma_y}{\sqrt{2}} \Big)  \otimes \Big(\dfrac{\sigma_x + \sigma_y}{\sqrt{2}} \Big)  \otimes \Big( \dfrac{\sigma_x + \sigma_y}{\sqrt{2}}\Big)$, 

$\bullet$ $\Big(\dfrac{\sigma_x - \sigma_y}{\sqrt{2}} \Big)  \otimes \Big(\dfrac{\sigma_x - \sigma_y}{\sqrt{2}} \Big) \otimes \Big(\dfrac{\sigma_x - \sigma_y}{\sqrt{2}} \Big)$.

\section{Single sided unsharp version of the witness operator $\mathcal{W}_W$}\label{appendix3}
Using the substitution $\langle \sigma_{x_i} \otimes \sigma_{y_j} \otimes \sigma^{\lambda_m}_{z^m_k} \rangle  \rightarrow \lambda_m \langle \sigma_{x_i} \otimes \sigma_{y_j} \otimes \sigma_{z^m_k} \rangle$ in the case of a general $\lambda_m$ \cite{bera},  the decomposition (\ref{decw}) of the genuine entanglement witness operator $\mathcal{W}_W$ can be written as,
\begin{align}
&\mathcal{W}_W^{\lambda_m}  =  \frac{1}{24} \Big(13 \, \mathbb{I} \otimes \mathbb{I} \otimes \mathbb{I} + 3 \, \sigma_z \otimes \mathbb{I} \otimes \mathbb{I} + 3 \, \mathbb{I}\otimes \sigma_z \otimes \mathbb{I} + 3 \, \mathbb{I} \otimes \mathbb{I}\otimes \lambda_m \, \sigma_z + 5 \, \sigma_z \otimes \sigma_z \otimes \mathbb{I} + 5 \, \sigma_z \otimes \mathbb{I}\otimes \lambda_m \, \sigma_z \nonumber \\
& + 5 \, \mathbb{I} \otimes \sigma_z \otimes \lambda_m \, \sigma_z + 7 \, \sigma_z\otimes \sigma_z \otimes \lambda_m \, \sigma_z - \mathbb{I} \otimes \mathbb{I} \otimes \lambda_m \, (\sigma_z + \sigma_x)  -  \mathbb{I}  \otimes (\sigma_z + \sigma_x) \otimes \mathbb{I} - (\sigma_z + \sigma_x) \otimes \mathbb{I} \otimes \mathbb{I} \nonumber \\
& - \mathbb{I} \otimes (\sigma_z + \sigma_x) \otimes \lambda_m \, (\sigma_z + \sigma_x)   - (\sigma_z + \sigma_x) \otimes \mathbb{I} \otimes \lambda_m \, (\sigma_z + \sigma_x) - (\sigma_z + \sigma_x)  \otimes (\sigma_z + \sigma_x) \otimes \mathbb{I}  \nonumber \\
&- (\sigma_z + \sigma_x) \otimes(\sigma_z + \sigma_x) \otimes \lambda_m \,(\sigma_z + \sigma_x)  - \mathbb{I} \otimes \mathbb{I} \otimes \lambda_m \, (\sigma_z - \sigma_x)  -  \mathbb{I}  \otimes (\sigma_z - \sigma_x) \otimes \mathbb{I} - (\sigma_z - \sigma_x) \otimes \mathbb{I} \otimes \mathbb{I} \nonumber \\
&- \mathbb{I} \otimes (\sigma_z - \sigma_x) \otimes \lambda_m \, (\sigma_z - \sigma_x)  - (\sigma_z - \sigma_x) \otimes \mathbb{I} \otimes \lambda_m \, (\sigma_z - \sigma_x) - (\sigma_z - \sigma_x)  \otimes (\sigma_z - \sigma_x) \otimes \mathbb{I} \nonumber \\
&- (\sigma_z - \sigma_x) \otimes(\sigma_z - \sigma_x) \otimes \lambda_m \, (\sigma_z - \sigma_x)  - \mathbb{I} \otimes \mathbb{I} \otimes \lambda_m \, (\sigma_z + \sigma_y)  -  \mathbb{I}  \otimes (\sigma_z + \sigma_y) \otimes \mathbb{I} - (\sigma_z + \sigma_y) \otimes \mathbb{I} \otimes \mathbb{I} \nonumber \\
&- \mathbb{I} \otimes (\sigma_z + \sigma_y) \otimes \lambda_m \, (\sigma_z + \sigma_y)  - (\sigma_z + \sigma_y) \otimes \mathbb{I} \otimes \lambda_m \, (\sigma_z + \sigma_y) - (\sigma_z + \sigma_y)  \otimes (\sigma_z + \sigma_y) \otimes \mathbb{I} \nonumber \\
& -(\sigma_z + \sigma_y) \otimes (\sigma_z + \sigma_y) \otimes \lambda_m \, (\sigma_z + \sigma_y)  - \mathbb{I} \otimes \mathbb{I} \otimes \lambda_m \, (\sigma_z - \sigma_y)  -  \mathbb{I}  \otimes (\sigma_z - \sigma_y) \otimes \mathbb{I} - (\sigma_z - \sigma_y) \otimes \mathbb{I} \otimes \mathbb{I} \nonumber \\
&- \mathbb{I} \otimes (\sigma_z - \sigma_y) \otimes \lambda_m \, (\sigma_z - \sigma_y)  - (\sigma_z - \sigma_y) \otimes \mathbb{I} \otimes \lambda_m \, (\sigma_z - \sigma_y) - (\sigma_z - \sigma_y)  \otimes (\sigma_z - \sigma_y) \otimes \mathbb{I} \nonumber \\
&-(\sigma_z - \sigma_y)  \otimes (\sigma_z - \sigma_y) \otimes \lambda_m \, (\sigma_z - \sigma_y) \Big).
\end{align}
\label{decwl}
Now, since we have Tr$[\mathcal{W}_W \rho_{BS}] \geq 0$ $\forall$ $\rho_{BS}$ $\in$ $ \mathcal{BS}$ (where $\mathcal{BS}$ is the set of all bi-seperable states) and $0 < \lambda_m \leq 1$, we can write the following:
\begin{align}\label{Wlambda}
&\mbox{Tr}[\mathcal{W}_W^{\lambda_m} \rho_{BS}] \nonumber \\
& = \lambda_m \mbox{Tr}[\mathcal{W}_W \rho_{BS}] + \frac{1}{24}(1-\lambda_m) \Big(13  + 3 \, \mbox{Tr}[ \rho_{BS}  (\sigma_z \otimes \mathbb{I} \otimes \mathbb{I})  ] + 3 \, \mbox{Tr}[ \rho_{BS}  ( \mathbb{I} \otimes \sigma_z \otimes \mathbb{I}  )]  + 5 \, \mbox{Tr}[ \rho_{BS}  ( \sigma_z \otimes \sigma_z \otimes \mathbb{I}  )]  \nonumber \\
&- \mbox{Tr}[ \rho_{BS}  ( \mathbb{I}  \otimes (\sigma_z + \sigma_x) \otimes \mathbb{I} )]  - \mbox{Tr}[ \rho_{BS}  ( (\sigma_z + \sigma_x) \otimes \mathbb{I} \otimes \mathbb{I}  )]   - \mbox{Tr}[ \rho_{BS}  ( (\sigma_z + \sigma_x)  \otimes (\sigma_z + \sigma_x) \otimes \mathbb{I}  )] \nonumber \\
& - \mbox{Tr}[ \rho_{BS}  (   \mathbb{I}  \otimes (\sigma_z - \sigma_x) \otimes \mathbb{I} )] - \mbox{Tr}[ \rho_{BS}  ( (\sigma_z - \sigma_x) \otimes \mathbb{I} \otimes \mathbb{I}  )]  - \mbox{Tr}[ \rho_{BS}  ( (\sigma_z - \sigma_x)  \otimes (\sigma_z - \sigma_x) \otimes \mathbb{I}  )]  \nonumber \\
& - \mbox{Tr}[ \rho_{BS}  (  \mathbb{I}  \otimes (\sigma_z + \sigma_y) \otimes \mathbb{I}  )]  - \mbox{Tr}[ \rho_{BS}  ( (\sigma_z + \sigma_y) \otimes \mathbb{I} \otimes \mathbb{I} )]    - \mbox{Tr}[ \rho_{BS}  ( (\sigma_z + \sigma_y)  \otimes (\sigma_z + \sigma_y) \otimes \mathbb{I}  )]   \nonumber \\
& - \mbox{Tr}[ \rho_{BS}  ( (\sigma_z - \sigma_y) \otimes \mathbb{I} \otimes \mathbb{I}  )] - \mbox{Tr}[ \rho_{BS}  ( \mathbb{I}  \otimes (\sigma_z - \sigma_y) \otimes \mathbb{I}  )] - \mbox{Tr}[ \rho_{BS}  ((\sigma_z - \sigma_y)  \otimes (\sigma_z - \sigma_y) \otimes \mathbb{I}  )]  \Big) \nonumber \\
&=\lambda_m \mbox{Tr}[\mathcal{W}_W \rho_{BS}] + \frac{1}{24}(1-\lambda_m) \Big(13  -\langle \sigma_z \otimes \mathbb{I} \otimes \mathbb{I} \rangle-\langle  \mathbb{I} \otimes \sigma_z \otimes \mathbb{I} \rangle + \langle \sigma_z \otimes \sigma_z \otimes \mathbb{I} \rangle - 2\langle \sigma_x \otimes \sigma_x \otimes \mathbb{I} \rangle- 2\langle \sigma_y \otimes \sigma_y \otimes \mathbb{I} \rangle \Big)\nonumber \\
& \geq \lambda_m \mbox{Tr}[\mathcal{W}_W \rho_{BS}] + \frac{1}{4}(1-\lambda_m)  \geq 0  \hspace{0.7cm} \forall \hspace{0.15cm} \rho_{BS} \hspace{0.15cm} \in \hspace{0.15cm} \mathcal{BS}.
\end{align}
The first inequality in the last line of \eqref{Wlambda} is obtained by minimizing all the expectation values. 

\section{Single sided unsharp version of the witness operator $\mathcal{W}_{GHZ}$}\label{appendix4}
After the substitution $\langle \sigma_{x_i} \otimes \sigma_{y_j} \otimes \sigma^{\lambda_m}_{z^m_k} \rangle  \rightarrow \lambda_m \langle \sigma_{x_i} \otimes \sigma_{y_j} \otimes \sigma_{z^m_k} \rangle$, the decomposition (\ref{decghz}) of the witness operator $\mathcal{W}_{GHZ}$ takes the following form, 
\begin{align*}
\mathcal{W}^{\lambda_m}_{GHZ} &= \frac{1}{8} \Big(3 \, \mathbb{I} \otimes \mathbb{I} \otimes \mathbb{I} - \mathbb{I} \otimes \sigma_z \otimes \lambda_m \, \sigma_z - \sigma_z \otimes \mathbb{I} \otimes \lambda_m \, \sigma_z -  \sigma_z \otimes \sigma_z \otimes \mathbb{I} - 2 \, \sigma_x \otimes \sigma_x \otimes \lambda_m \, \sigma_x \nonumber \\
& +  \frac{1}{2}(\sigma_x + \sigma_y) \otimes (\sigma_x + \sigma_y) \otimes \lambda_m \, (\sigma_x + \sigma_y) + \frac{1}{2}(\sigma_x - \sigma_y) \otimes (\sigma_x - \sigma_y) \otimes \lambda_m \, (\sigma_x - \sigma_y) \Big).
\label{decghzl}
\end{align*}

Now, since we have Tr$[\mathcal{W}_{GHZ} \rho_{BS}] \geq 0$ $\forall$ $\rho_{BS}$ $\in$ $ \mathcal{BS}$ and $0 < \lambda_m \leq 1$, we can write the following:
\begin{align}
&\mbox{Tr}[\mathcal{W}_{GHZ}^{\lambda_m} \rho_{BS}]\ \nonumber \\
& = \lambda_m \mbox{Tr}[\mathcal{W}_{GHZ} \rho_{BS}]   + \frac{1}{8}  (1-\lambda_m)\Big(3 -  \langle \sigma_z \otimes \sigma_z \otimes \mathbb{I} \rangle \Big) \nonumber \\
& \geq \lambda_m \mbox{Tr}[\mathcal{W}_{GHZ} \rho_{BS}] + \frac{1}{4}(1-\lambda_m) \nonumber \\
& \geq 0  \hspace{0.7cm} \forall \hspace{0.15cm} \rho_{BS} \hspace{0.15cm} \in \hspace{0.15cm} \mathcal{BS}.
\end{align}

\end{widetext}

\end{document}